\documentclass[
    ,final            
  ]
  {aipproc}

\layoutstyle{6x9}

\def\simgt{\lower 2pt \hbox{$\, \buildrel {\scriptstyle >}\over {\scriptstyle \sim}\,$}}
\def\simlt{\lower 2pt \hbox{$\, \buildrel {\scriptstyle <}\over {\scriptstyle \sim}\,$}}

\def\astroe2{{\it {\it Astro-E2}\/}}

\def\chandra{{\it Chandra\/}}
\def\conx{{\it Constellation-X\/}}

\def\fuse{{\it FUSE\/}}
\def\hst{{\it {\it HST}\/}}
\def\iue{{\it IUE\/}}

\def\xeus{{\it XEUS\/}}
\def\xmm{{\it XMM-Newton\/}}

\begin{document}

\title{Grating X-ray Spectroscopy of High-Velocity Outflows from Active Galaxies}

\author{W.N. Brandt}{
  address={Department of Astronomy \& Astrophysics, 525 Davey Laboratory, 
The Pennsylvania State University, University Park, Pennsylvania 16802, USA}
}

\author{S. Kaspi}{
  address={School of Physics and Astronomy and the Wise Observatory, The Raymond and Beverly Sackler
Faculty of Exact Sciences, Tel-Aviv University, Tel-Aviv 69978, Israel}
}

\begin{abstract}
X-ray absorption and emission lines now serve as powerful diagnostics
of the outflows from active galaxies. Detailed X-ray line studies of 
outflows have recently been enabled for a significant number of 
active galaxies via the grating spectrometers on \chandra\ and
\xmm. We will review some of the recent X-ray findings on active
galaxy outflows from an observational perspective. We also describe
some future prospects. 

X-ray absorption lines from H-like and He-like ions of C, N, O, Ne, 
Mg, Al, Si, and S are often seen. A wide range of ionization parameter
appears to be present in the absorbing material, and inner-shell 
absorption lines from lower ionization ions, Fe~L-shell lines, and 
Fe~M-shell lines have also been seen. The X-ray absorption lines are 
typically blueshifted relative to the systemic velocity by a few
hundred km~s$^{-1}$, and they often appear kinematically
consistent with UV absorption lines of C~{\sc iv}, N~{\sc v}, and 
H~{\sc i}. The X-ray absorption lines can have complex profiles with 
multiple kinematic components present as well as filling of the 
absorption lines by emission-line photons. A key remaining uncertainty 
is the characteristic radial location of the outflowing gas; only after 
this quantity is determined will it be possible to calculate reliably 
the amount of outflowing gas and the kinetic luminosity of the outflow. 
\end{abstract}

\maketitle



\section{Introduction}

Outflows are observed to be ubiquitous in active galactic nuclei (AGN), being
seen in objects spanning a range of $\sim 10,000$ in luminosity. They have been
studied in the most detail via observations of ultraviolet (UV) resonance lines
from moderately ionized gas. In luminous Broad Absorption Line quasars (BALQSOs), 
outflows are observed to reach velocities up to a few $10^4$~km~s$^{-1}$, and they 
subtend $\approx$~10--30\% of the sky as viewed from the central source. In 
lower luminosity Seyfert galaxies, outflows are observed $\simgt 50$\% of the 
time although they have velocities up to only $\approx 10^3$~km~s$^{-1}$. 
These outflows are a major component of the nuclear environment, and they 
may carry a significant fraction of the accretion power. They may also be 
important in regulating the growth of the black hole and its host galaxy
(e.g., Silk \& Rees 1998; Fabian 1999) as well as in injecting matter, energy, 
and magnetic fields into the intergalactic medium (e.g., Turnshek 1988; 
Wu, Fabian, \& Nulsen 2000; Furlanetto \& Loeb 2001; Elvis et~al. 2002). The observed
outflows are photoionized by the radiation from the central source, and 
they are probably driven by radiation pressure. Despite their ubiquity and 
importance, their physical location and origin in the AGN system remain 
unclear; outflows may arise from winds driven off the surface of an 
accretion disk (e.g., Murray et~al. 1995; Proga, Stone, \& Kallman 2000; Elvis 2000), 
a dusty torus (e.g., Voit, Weymann, \& Korista 1993; Krolik \& Kriss 1995), or 
perhaps stars in the nucleus (e.g., Scoville \& Norman 1995; Netzer 1996). 

\begin{figure}[t!]
  \includegraphics[height=5 in,width=4.5 in]{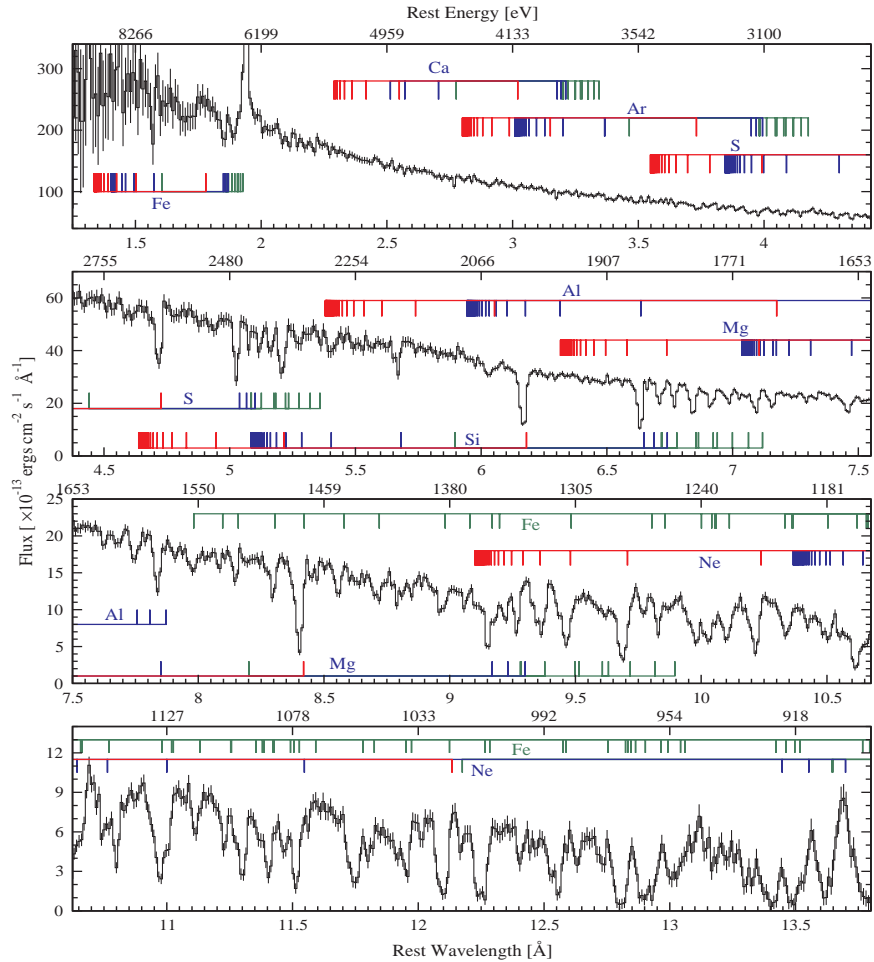}
  \caption{Part of the 10.4-day \chandra\ HETGS spectrum of the Seyfert 
galaxy NGC~3783. Marked are the large number of detected absorption 
lines as well as several emission lines. The lines are marked at
their expected wavelengths in the rest frame of NGC~3783; the 
blueshifts of the absorption lines are noticeable. In total, more
than 140 spectral features are detected in the X-ray spectrum of
NGC~3783. Adapted from Kaspi et~al. (2002).}
\end{figure}

Ionized absorption in the X-ray band has been intensively studied in 
bright, low-redshift AGN for over a decade (e.g., the ``warm absorbers'' in 
Seyfert galaxies; Reynolds 1997; George et~al. 1998). The luminous X-ray 
source in the nucleus acts as a ``flashlight'' allowing observers to 
``X-ray'' material along the line of sight. However, prior to the launches 
of \chandra\ and \xmm, such investigations were limited by a lack of 
spectral resolution. Over the past three years, the X-ray grating 
spectrometers on these two missions\footnote{The relevant instruments are 
the \chandra\ High-Energy Transmission Grating Spectrometer 
(HETGS; C.R. Canizares et~al., in preparation), 
the \chandra\ Low-Energy Transmission Grating Spectrometer
(LETGS; Brinkman et~al. 1997), and
the \xmm\ Reflection Grating Spectrometer
(RGS; den~Herder et~al. 2001).} 
have enlarged the number of spectral 
features available for study by a factor of $\sim 50$
(see Fig.~1; from 2--3 to more than 140). They have improved the 
velocity resolution available to observers from $\sim 15,000$~km~s$^{-1}$ 
to $\sim 400$~km~s$^{-1}$. They have thereby provided qualitatively 
new information on the physical conditions, kinematics, and 
geometry/location of the absorbing material. 

At present, efficient grating X-ray spectroscopy is possible only for
$\approx 20$ bright, low-redshift (mainly $z<0.1$) AGN, mainly Seyfert 
galaxies. Even for these, the required observation lengths are 
typically $\simgt$~1--2~days; the spectrum of NGC~3783 shown in Fig.~1 
required a 10.4-day exposure with \chandra. Of necessity, the discussion 
below applies predominantly to this fairly small sample of objects. 
Highly luminous and distant quasars, for example, may 
have significantly different X-ray absorption properties. 
Furthermore, the discussion below will be closely tied to the X-ray 
observations, without detailed descriptions of theoretical models. For 
further information on theoretical models, the reader should consult 
one of the current reviews (e.g., Netzer 2001; Krolik 2002; and 
references therein). 


\begin{figure}[t!]
  \includegraphics[height=2.5 in,width=5.7 in]{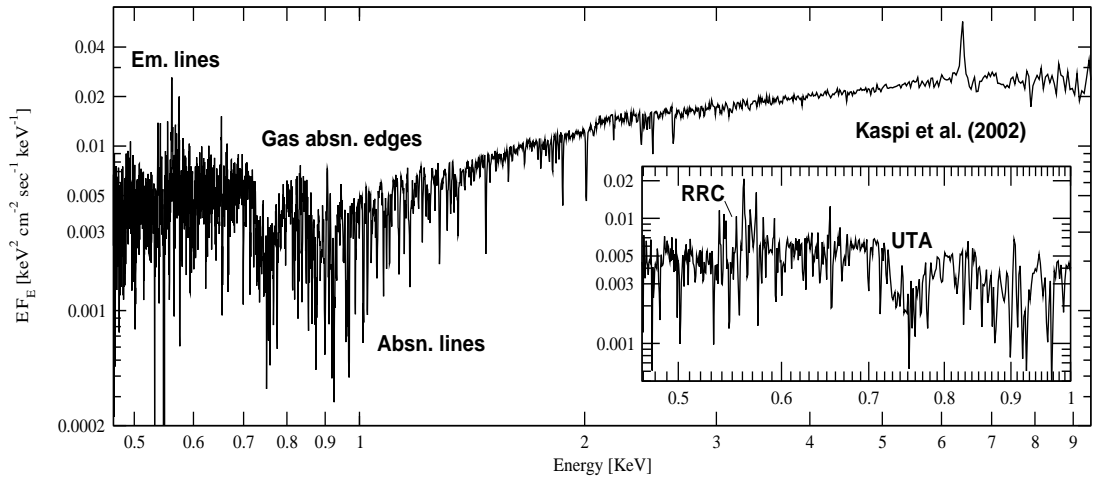}
  \caption{The 10.4-day \chandra\ HETGS spectrum of the Seyfert 
galaxy NGC~3783 shown in $EF_{\rm E}$ vs. energy format. The
insert focuses on the spectrum below 1~keV. The different types
of observed spectral features are labeled. Adapted from 
Kaspi et~al. (2002).}
\end{figure}

\section{Some Key Results from X-ray Gratings Studies}

The spectral features seen in the current X-ray gratings studies include
absorption lines [sometimes in unresolved transition arrays (UTAs)], 
emission lines, 
absorption edges (from gas and perhaps dust), and
radiative recombination continua (RRCs).\footnote{See the 
conference papers by E.~Behar and T.~Kallman for further discussion
of some of these features.} The X-ray absorption lines are the
most numerous features and are from H-like and He-like ions of C, N, 
O, Ne, Mg, Al, Si, and S. Inner-shell absorption lines from lower 
ionization ions, Fe~L-shell lines, and Fe~M-shell lines are also 
seen. Fig.~2 illustrates the observed features
in the spectrum of NGC~3783. Taken together, these features are 
sensitive to a wide range of ionization parameter and column 
density, and they also provide useful temperature and density 
diagnostics. Furthermore, the ionizing continuum is directly
visible (unlike the case for UV absorption lines,
where the ionizing continuum is displaced in wavelength), 
and X-rays are relatively immune to dust extinction effects 
that can hinder studies at other wavelengths. 

\begin{figure}[t!]
  \includegraphics[height=5 in,width=5.0 in]{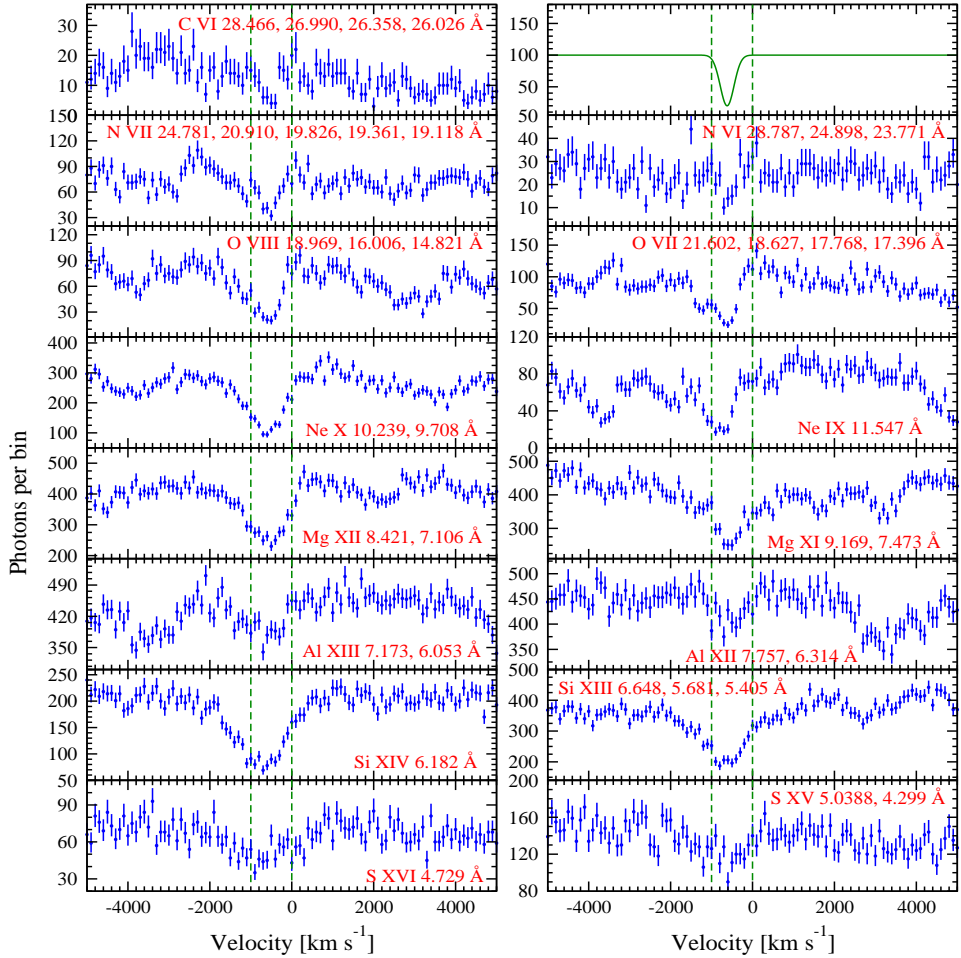}
  \caption{Velocity spectra showing co-added lines from different ions
in the 10.4-day \chandra\ HETGS spectrum of the Seyfert galaxy NGC~3783. 
H-like ions are shown on the left, and He-like ions are shown on the right. 
The bin size is 100~km~s$^{-1}$, and vertical dashed lines are given at
velocities of 0~km~s$^{-1}$ and $-1000$~km~s$^{-1}$ to guide the eye. 
In the uppermost right panel we show a Gaussian absorption line 
representing the line response function of the instrument at 17.396~\AA\ 
(the FWHM is 397~km~s$^{-1}$); this is the poorest line response function 
applicable to the co-added velocity spectrum of O~{\sc vii}. Note the 
asymmetry of the O~{\sc vii} lines that is apparently from an additional 
absorption system. Adapted from Kaspi et~al. (2002).}
\end{figure}

\subsection{Physical Conditions in Outflows}

Gratings observations have clearly established that a uniform, single-component
model for the X-ray absorbing material in the outflow is too simple. 
Outflows contain gas with a wide range of ionization parameter. Fig.~3, 
for example, shows X-ray absorption lines from the wide range of ions 
observed in NGC~3783. It is not possible to fit both the low-ionization
and high-ionization absorption lines simultaneously with a single ``zone'' 
of photoionized gas (e.g., Kaspi et~al. 2002; Blustin et~al. 2002). 
Similar results have been found for several other AGN 
(e.g., Sako et~al. 2001; Kaastra et~al. 2002). Current 
models for the outflow usually assume the absorbing 
gas is in photoionization equilibrium. This is a 
plausible assumption for some AGN, such as NGC~3783, that exhibit fairly 
slow and small-amplitude X-ray variability. However, AGN with more rapid 
and large-amplitude variability (e.g., NGC~4051 and other 
``Narrow-Line Seyfert~1'' galaxies) may contain non-equilibrium 
X-ray absorbers (e.g., Nicastro et~al. 1999; Collinge et~al. 2001). 

\begin{figure}[t!]
  \includegraphics[height=4 in,width=3.5 in]{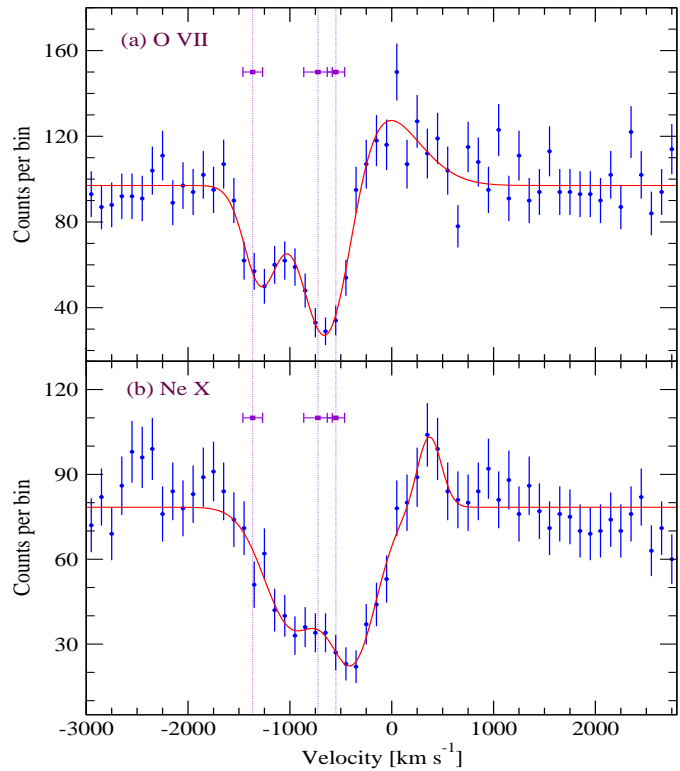}
  \caption{Velocity spectra, binned at 100~km~s$^{-1}$ resolution, 
of (a) four strong lines of O~{\sc vii} and (b) two strong lines
of Ne~{\sc x} in the 10.4-day \chandra\ HETGS spectrum of the 
Seyfert galaxy NGC~3783. A three-Gaussian fit to each spectrum
individually is overplotted on the data points (two Gaussians
are in absorption, and one is in emission). Two absorption 
systems are clearly detected in O~{\sc vii} and probably
exist in Ne~{\sc x} as well, although there are statistically
significant differences between the O~{\sc vii} and Ne~{\sc x}
absorption-line profiles. The vertical lines show the
velocity shifts of the observed UV absorption systems; the 
squares with horizontal error bars show the FWHMs of these 
systems. Adapted from Kaspi et~al. (2002).} 
\end{figure}

Early photoionization modeling of the ionized absorbers in Seyfert 
galaxies suggested that they have temperatures $(T)$ of a few 
$10^5$~K. The gratings data now confirm the expected temperatures. 
In NGC~3783, for example, modeling of the O~{\sc vii} and N~{\sc vi} RRCs 
indicate $T\simgt 6\times 10^4$~K, and constraints from He-like triplet 
emission lines require $T\simlt 10^6$~K (Kaspi et~al. 2002). 

Column densities for some of the stronger absorption lines detected
from AGN outflows can be estimated via ``curve of growth'' analyses. 
However, in such analyses, it is essential to avoid lines that are
saturated. The identification of saturated lines can be difficult, 
since they need not appear ``black'' (i.e., drop to zero intensity) 
due to the presence of nuclear X-ray scattering or multiple, unresolved 
line components (compare Hamann 1998 and Arav et~al. 1999). 
Lines representing transitions to high atomic levels 
are the least likely to be affected by saturation (due to their 
lower oscillator strengths), and analyses of such lines indicate 
O~{\sc vii} and O~{\sc viii} column densities of a few
$10^{18}$~cm$^{-2}$. The corresponding total 
hydrogen column densities, assuming solar abundances and a reasonable 
ionization correction, range from a few $10^{21}$~cm$^{-2}$ to 
$\approx 10^{22}$~cm$^{-2}$. Considering the statistical and systematic 
uncertainties currently present in curve of growth analyses, solar 
abundances usually appear consistent with the data. 
Some AGN outflows may also contain significant column densities
of metals in the form of dust grains; X-ray gratings studies
offer the exciting possibility of detecting these grains directly
and measuring their chemical composition
(e.g., Lee et~al. 2001). 

\begin{figure}[t!]
  \includegraphics[height=4.5 in,width=3.0 in,angle=-90]{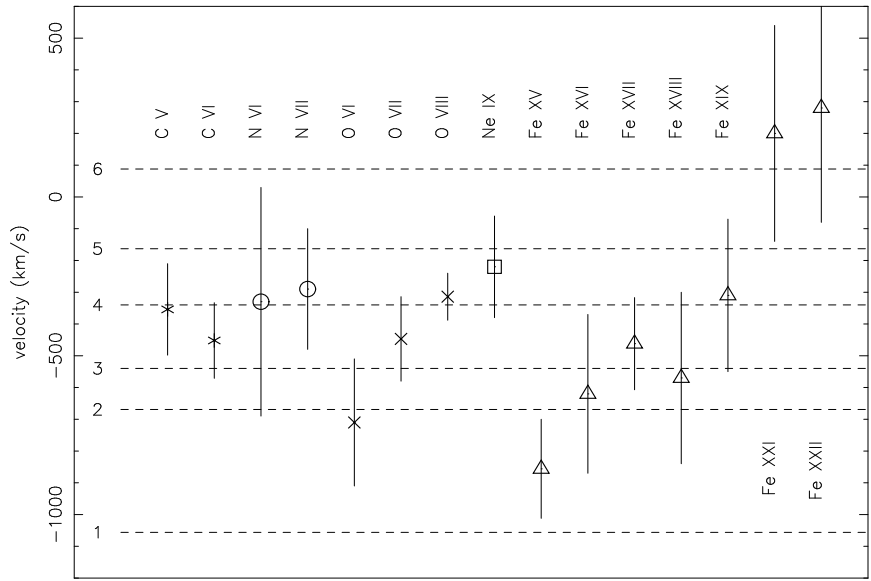}
  \caption{Average velocities (with respect to the AGN rest frame) for 
the X-ray absorption lines of different ions in the \chandra\ LETGS 
spectrum of the Seyfert galaxy NGC~5548. The dashed horizontal lines
show the velocities of the six absorption components identified in 
the UV. Note the tendency for less ionized iron ions to have 
larger outflow velocities. Adapted from Kaastra et~al. (2002).}
\end{figure}

\subsection{Kinematics of Outflows}

The most important, basic result on kinematics from X-ray gratings
studies is that the X-ray absorbing gas is generally in a state of
outflow with a bulk velocity of a few hundred km~s$^{-1}$ (as
derived from the blueshifts of X-ray absorption lines). Prior to 
\chandra\ and \xmm, it was not possible to speak reliably about the
X-ray absorber as an outflow! In several cases, it has also now 
been possible to resolve the X-ray absorption lines. X-ray absorbing
outflows can have velocity dispersions comparable to their bulk 
velocities, perhaps due to acceleration of the outflow or turbulence. 
Furthermore, in a few cases, multiple kinematic components in a single
ion have been discovered (e.g., see Fig.~4). 
As illustrated in Fig.~4 and Fig.~5, different ions can have
different kinematic properties; ionization level and kinematics 
are therefore connected. 
It is important to note that there may be systematic errors
in velocity measurements derived from X-ray absorption lines. 
For example, some X-ray lines show ``P~Cygni'' profiles where
the emission line partially fills the red side of the 
absorption line. If this (geometry dependent) effect is not 
properly modeled, velocity measurements can be biased to be
too large (because the effective centroid of the fitted 
absorption line is shifted blueward). This effect appears
to be present in the 10.4-day \chandra\ HETGS spectrum of 
NGC~3783 (Kaspi et~al. 2002; although alternative interpretations 
are possible), and it may well be present but difficult to correct 
in X-ray gratings spectra with lower signal-to-noise. 

The outflow velocities measured from X-ray absorption lines
often agree with those measured from UV absorption lines such 
as C~{\sc iv}, N~{\sc v}, and H~{\sc i} (e.g., see 
Fig.~4 and Fig.~5), supporting the general existence of a 
connection between these two types of absorption 
(e.g., Mathur, Elvis, \& Wilkes 1995; Crenshaw et~al. 1999). 
The exact nature of this connection, however, depends upon 
details of the modeling and remains debated 
(e.g., Crenshaw \& Kraemer 1999; 
Arav, Korista, \& de~Kool 2002; and references therein). 

\subsection{Geometry and Radial Location of Outflows}

From the relative strengths of the emission and absorption lines in 
the AGN with X-ray gratings spectra, the global covering factor of 
the X-ray absorbing outflow appears to be large.\footnote{The global
covering factor is the fraction of the sky, as seen from the
central source, covered by the X-ray absorbing outflow.} The outflow 
covers $\simgt 50$\% of the sky that is not already covered by the 
torus of AGN unification schemes. This direct constraint 
on the global covering factor agrees well with the indirect 
constraint derived from counting the fraction of local Seyfert~1 
galaxies with ionized X-ray absorbers. The line-of-sight covering
factor also appears to be large in at least a few AGN; it can be 
constrained by measuring the extent to which saturated lines
appear black.\footnote{The line-of-sight covering factor is the 
fraction of the line-of-sight to the central source covered by 
the X-ray absorbing outflow. In reality, there may be multiple 
lines of sight to the central source if it has physical extent
or if X-ray scattering is present.} In NGC~3783, for example, 
the Fe~{\sc xx} lines near 12.8~\AA\ (see Fig.~1) limit any 
electron-scattered X-ray contribution to be $\simlt 15$\% 
(Kaspi et~al. 2002). 

The most important remaining uncertainty about X-ray absorbing
outflows is their radial location; this key quantity is not directly 
constrained by a single observation of an AGN. Possible radial
locations range from \hbox{$\sim 10^{16}$~cm} (e.g., an accretion disk
wind) to \hbox{$\sim 10^{18}$~cm} (e.g., a torus wind), and there may 
be absorbing material across this entire range of radii. Knowledge of the 
radial location is essential for determining the mass outflow
rate, the kinetic energy of the outflow, and the overall 
importance of the outflow in the AGN system. One physically 
appealing possibility is that the material in the X-ray absorbing 
outflow is the same material that scatters radiation to the 
observer in Seyfert~2 galaxies (e.g., Krolik \& Kriss 1995; 
Krolik \& Kriss 2001), but this connection cannot be 
firmly established until the radial location of the
X-ray absorbing outflow is known. Variability studies combined
with improved density diagnostics may allow the radial location
to be determined (e.g., Krolik \& Kriss 2001; Netzer et~al. 2002), 
but this will require expensive \chandra\ and \xmm\ observations 
at multiple epochs. 


\section{Possible Future Directions}

\subsection{The Need for Better X-ray Spectral Resolution}

While the new data from \chandra\ and \xmm\ represent an {\it enormous\/} 
advance, it is likely that X-ray spectroscopy of AGN outflows is still
limited in some fundamental ways. One possible problem is illustrated in 
Fig.~6, where \chandra, \hst, and \iue\ spectra of the Seyfert
galaxy NGC~4051 are compared. Given the results from UV observations, 
it appears at least plausible that the ionized X-ray absorbers in AGN
have significant velocity structure that cannot be resolved with current 
X-ray instruments. The velocity structure currently apparent in the X-ray 
spectra (e.g., Fig.~4 and Fig.~5) may just be the ``tip of the iceberg.''
An optimist might argue that the high-temperature X-ray absorbing 
material will tend to be more volume filling than the lower temperature
UV absorbing material, and therefore that it will be less clumpy. 
However, \fuse\ spectra of lines from the high-ionization ion O~{\sc vi} 
still show velocity structure finer than can be resolved with current 
X-ray instruments (e.g., Gabel et~al. 2002). 

\begin{figure}[t!]
  \includegraphics[height=3.5 in]{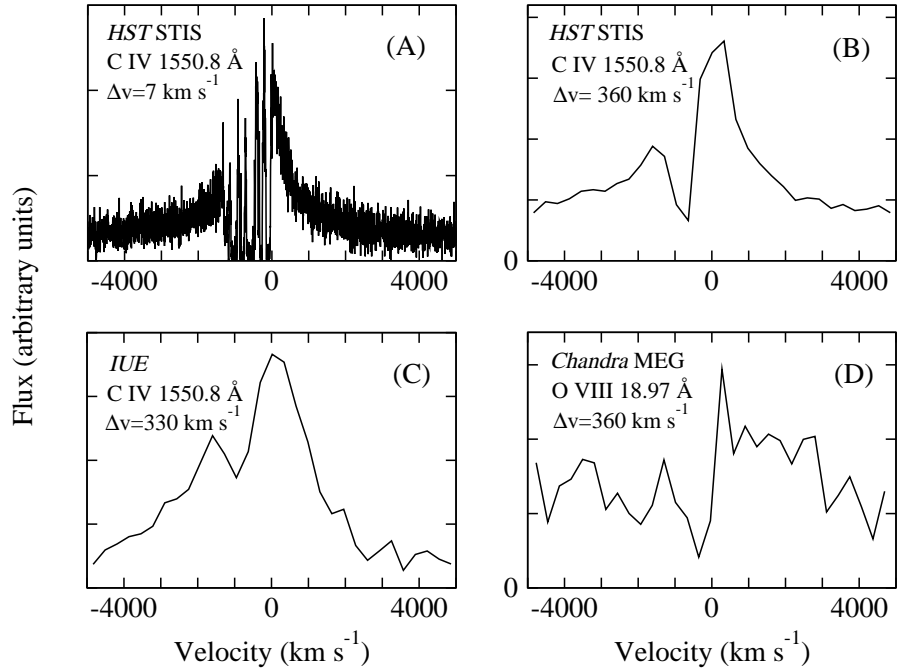}
  \caption{UV and X-ray absorption systems in the Seyfert 
galaxy NGC~4051 as seen by 
the \hst\ Space Telescope Imaging Spectrograph (STIS; a and b),
\iue\ (c; mean of 1978--1988 and 1994 epochs), and
the \chandra\ HETGS (d; the MEG is part of the HETGS). 
The \chandra\ and \hst\ observations were performed 
simultaneously in 2000. The approximate velocity resolution 
is listed for each panel. Panel (b) represents the STIS 
C~{\sc iv} line binned to the velocity resolution of the HETGS. 
Comparison of the panels stresses the point that the ionized 
X-ray absorber may be subdivided into further systems 
that cannot be resolved with current X-ray instruments. 
Note the similarity between panels (b) and (d), even though
the high resolution in panel (a) reveals the C~{\sc iv} 
absorption to be extremely complex with at least eight 
distinct kinematic components. Adapted from 
Collinge et~al. (2001).}
\end{figure}

If the X-ray absorption lines indeed possess significant unresolved
structure, this will lead to systematic errors when attempting to 
derive column densities, ionization levels, and other physical 
parameters. Some line components may be saturated even though the
unresolved average ``line'' does not appear black. A 
velocity resolution of $\sim 100$~km~s$^{-1}$ 
or better will probably be required to resolve the X-ray lines. Such 
a resolution would also be valuable for addressing numerous other
astrophysical issues (e.g., Elvis 2001). To our knowledge, no X-ray 
missions with the requisite resolution are currently scheduled for 
launch. 

\subsection{The Need for Higher Throughput}

As mentioned above, efficient grating X-ray spectroscopy is presently
possible only for $\approx 20$ bright, low-redshift (mainly $z<0.1$) AGN. 
Photon starvation has limited our ability to 
perform grating X-ray spectroscopy of more luminous but more distant AGN,
such as BALQSOs. This is unfortunate, since the outflows in these objects
are faster and probably more powerful than those in local AGN
(e.g., see Laor \& Brandt 2002 and references therein). The 
current X-ray spectra of BALQSOs, obtained with Charged Coupled Device (CCD)  
detectors, show that heavy X-ray absorption is often present
(e.g., Green et~al. 2001; Gallagher et~al. 2002). In most cases, 
however, the dynamical state of the X-ray absorbing gas is unknown. 
One notable exception appears to be the gravitationally lensed BALQSO 
APM~08279+5255 at $z=3.91$. Chartas et~al. (2002) have recently claimed 
the detection of X-ray BALs from iron~K$\alpha$ that imply outflow velocities 
of $\approx$~(0.2--0.4)$c$ (Fig.~7; also see Hasinger, Schartel, \& Komossa 2002 
for additional X-ray observations and an alternative interpretation). 

\begin{figure}[t!]
  \includegraphics[height=3.5 in,angle=-90]{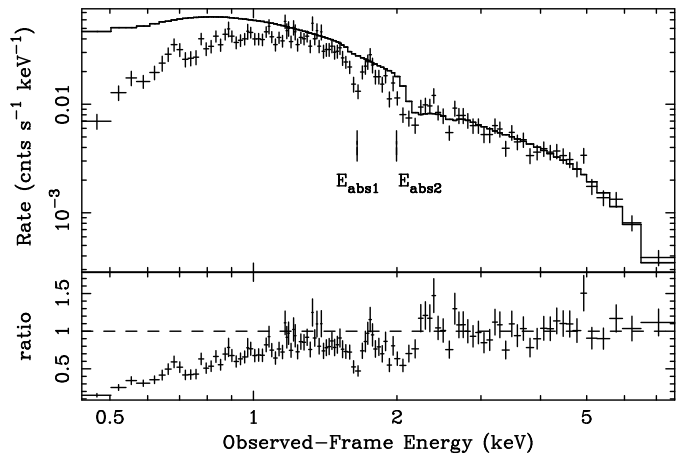}
  \caption{The observed-frame \chandra\ spectrum of the $z=3.91$
BALQSO APM~08279+5255. The spectrum has been fit above 2.2~keV
(10.8~keV in the rest frame) with a power-law plus Galactic
absorption model; the model has then been extrapolated to
lower energies to show the residuals. The lower panel shows
the ratio of the data to the model. Note the absorption at
low energies as well as the two apparent absorption lines
from 1.5--2.1~keV. Adapted from Chartas et~al. (2002).}
\end{figure}

High-throughput spectroscopy with enough resolution to determine
the velocities of the X-ray absorbers in BALQSOs should provide a 
qualitative advance in our understanding of BALQSO outflows. Without
basic kinematic information on the X-ray absorber, it is impossible
to determine the total mass outflow rate and the kinetic energy of
the outflow. Extremely high spectral resolution is probably not 
required for this science due to the high velocities observed
in the UV; a velocity resolution of $\sim 1000$~km~s$^{-1}$ should
be sufficient to allow major progress. Long \xmm\ observations of 
1--3 of the X-ray brightest BALQSOs can provide crucial information, 
but missions such as \conx\ and \xeus\ will be needed to make 
high-resolution X-ray spectroscopy of BALQSOs routine.  



\begin{theacknowledgments}
We thank all our collaborators, and we thank
the conference organizers for a stimulating conference. 
We gratefully acknowledge support from 
NASA LTSA grant NAG5-8107 (WNB, SK), 
CXC grant GO1-2103B (WNB, SK), 
CXC grant GO0-1041X (WNB), and
the Israel Science Foundation (SK). 
\end{theacknowledgments}


\section{References}

\footnotesize

\vskip 0.02 in 
\par\noindent\hangindent=10pt\hangafter=1
Arav N., Korista K.T., de~Kool M., Junkkarinen V.T., Begelman M.C.,
1999, 
`Hubble Space Telescope observations of the broad absorption line 
quasar PG~0946+301',
ApJ, 516, 27

\vskip 0.02 in 
\par\noindent\hangindent=10pt\hangafter=1
Arav N., Korista K.T., de~Kool M., 
2002, 
`On the column density of AGN outflows: The case of NGC~5548',
ApJ, 566, 699

\vskip 0.02 in 
\par\noindent\hangindent=10pt\hangafter=1
Blustin A.J., Branduardi-Raymont G., Behar E., Kaastra J.S., Kahn S.M., 
Page M.J., Sako M., Steenbrugge K.C., 
2002, 
`Multi-wavelength study of the Seyfert~1 galaxy NGC~3783 with \xmm', 
A\&A, in press (astro-ph/0206316)

\vskip 0.02 in 
\par\noindent\hangindent=10pt\hangafter=1
Brinkman A.C., et~al., 
1997, 
`Preliminary test results on spectral resolution of the Low-Energy
Transmission Grating Spectrometer on board AXAF',
Proc. SPIE, 3113, 181

\vskip 0.02 in 
\par\noindent\hangindent=10pt\hangafter=1
Chartas G., Brandt W.N., Gallagher S.C., Garmire G.P., 
2002, 
`\chandra\ detects relativistic broad absorption lines from APM~08279+5255', 
ApJ, in press (astro-ph/0207196) 

\vskip 0.02 in 
\par\noindent\hangindent=10pt\hangafter=1
Collinge M.J., Brandt W.N., Kaspi S., Crenshaw D.M., Elvis M., 
Kraemer S.B., Reynolds C.S., Sambruna R.M., Wills B.J., 
2001, 
`High-resolution X-ray and ultraviolet spectroscopy of the complex intrinsic
absorption in NGC~4051 with \chandra\ and the Hubble Space Telescope',
ApJ, 557, 2

\vskip 0.02 in 
\par\noindent\hangindent=10pt\hangafter=1
Crenshaw D.M., Kraemer S.B., 
1999, 
`Intrinsic absorption lines in the Seyfert~1 galaxy NGC~5548:
Ultraviolet echelle spectra from the Space Telescope Imaging
Spectrograph',
ApJ, 521, 572

\vskip 0.02 in 
\par\noindent\hangindent=10pt\hangafter=1
Crenshaw D.M., Kraemer S.B., Boggess A., Maran S.P., Mushotzky R.F., Wu C., 
1999, 
`Intrinsic absorption lines in Seyfert~1 galaxies. I. Ultraviolet 
spectra from the Hubble Space Telescope',
ApJ, 516, 750
 
\vskip 0.02 in 
\par\noindent\hangindent=10pt\hangafter=1
den~Herder J.W., et~al., 
2001, 
`The Reflection Grating Spectrometer on board \xmm', 
A\&A, 365, L7

\vskip 0.02 in 
\par\noindent\hangindent=10pt\hangafter=1
Elvis M., 
2000, 
`A structure for quasars', 
ApJ, 545, 63

\vskip 0.02 in 
\par\noindent\hangindent=10pt\hangafter=1
Elvis M., 
2001, 
`Thermal limit spectroscopy as a goal for X-ray astronomy', 
in Inoue H., Kunieda H., eds, 
New Century of X-ray Astronomy. 
ASP Press, San Francisco, p.~180

\vskip 0.02 in 
\par\noindent\hangindent=10pt\hangafter=1
Elvis M., Marengo M., Karovska M., 
2002, 
`Smoking quasars: A new source for cosmic dust', 
ApJ, 567, L107

\vskip 0.02 in 
\par\noindent\hangindent=10pt\hangafter=1
Fabian A.C., 
1999, 
`The obscured growth of massive black holes',
MNRAS, 308, L39

\vskip 0.02 in 
\par\noindent\hangindent=10pt\hangafter=1
Furlanetto S.R., Loeb A., 
2001, 
`Intergalactic magnetic fields from quasar outflows',
ApJ, 556, 619

\vskip 0.02 in 
\par\noindent\hangindent=10pt\hangafter=1
Gabel J.R., Crenshaw D.M., Kraemer S.B., et~al., 
2002, 
`The ionized gas and nuclear environment in NGC~3783: II. Averaged
HST STIS and FUSE spectra', 
ApJ, submitted

\vskip 0.05 in 
\par\noindent\hangindent=10pt\hangafter=1
Gallagher S.C., Brandt W.N., Chartas G., Garmire G.P., 
2002, 
`X-ray spectroscopy of quasi-stellar objects with broad ultraviolet absorption lines',
ApJ, 567, 37

\vskip 0.02 in 
\par\noindent\hangindent=10pt\hangafter=1
George I.M., Turner T.J., Netzer H., Nandra K., Mushotzky R.F., Yaqoob T., 
1998, 
`ASCA observations of Seyfert~1 galaxies. III. The evidence for absorption
and emission due to photoionized gas', 
ApJS, 114, 73

\vskip 0.05 in 
\par\noindent\hangindent=10pt\hangafter=1
Green P.J., Aldcroft T.L., Mathur S., Wilkes B.J., Elvis M., 
2001, 
`A \chandra\ survey of broad absorption line quasars',
ApJ, 558, 109

\vskip 0.02 in 
\par\noindent\hangindent=10pt\hangafter=1
Hamann F., 
1998, 
`Broad P~V absorption in the QSO PG~1254+047: Column densities, ionizations, 
and metal abundances in broad absorption line winds',
ApJ, 500, 798

\vskip 0.02 in 
\par\noindent\hangindent=10pt\hangafter=1
Hasinger G., Schartel N., Komossa S., 
2002, 
`Discovery of an ionized Fe~K edge in the $z=3.91$ broad absorption line 
quasar APM 08279+5255 with \xmm',
ApJ, 573, L77

\vskip 0.02 in 
\par\noindent\hangindent=10pt\hangafter=1
Kaastra J.S., Steenbrugge K.C., Raassen A.J.J., van~der~Meer R.L.J., Brinkman A.C., 
Liedahl D.A., Behar E., de~Rosa A., 
2002, 
`X-ray spectroscopy of NGC~5548', 
A\&A, 386, 427

\vskip 0.02 in 
\par\noindent\hangindent=10pt\hangafter=1
Kaspi S., Brandt W.N., George I.M., Netzer H., et~al., 
2002, 
`The ionized gas and nuclear environment in NGC~3783: I. Time-averaged
900~ks \chandra\ grating spectroscopy', 
ApJ, 574, 643

\vskip 0.02 in 
\par\noindent\hangindent=10pt\hangafter=1
Krolik J.H., Kriss G.A., 
1995, 
`Observable properties of X-ray-heated winds in active galactic
nuclei: Warm reflectors and warm absorbers', 
ApJ, 447, 512

\vskip 0.02 in 
\par\noindent\hangindent=10pt\hangafter=1
Krolik J.H., Kriss G.A., 
2001, 
`Warm absorbers in active galactic nuclei: A multitemperature wind', 
ApJ, 561, 684

\vskip 0.02 in 
\par\noindent\hangindent=10pt\hangafter=1
Krolik J.H., 
2002, 
`High-resolution X-ray spectroscopy and the nature of warm 
absorbers in AGN', 
in Boller Th., Komossa S., Kahn S., Kunieda H., eds, 
X-ray spectroscopy of AGN with \chandra\ and \xmm. 
MPE Press, Garching, in press (astro-ph/0204418)

\vskip 0.02 in 
\par\noindent\hangindent=10pt\hangafter=1
Laor A., Brandt W.N., 
2002, 
`The luminosity dependence of ultraviolet 
absorption in active galactic nuclei', 
ApJ, 569, 641

\vskip 0.02 in 
\par\noindent\hangindent=10pt\hangafter=1
Lee J.C., Ogle P.M., Canizares C.R., Marshall H.L., Schulz N.S., 
Morales R., Fabian A.C., Iwasawa K., 
2001, 
`Revealing the warm absorber in MCG--6--30--15 with the \chandra\ HETG',
ApJ, 554, L13

\vskip 0.02 in 
\par\noindent\hangindent=10pt\hangafter=1
Mathur S., Elvis M., Wilkes B.J., 
1995, 
`Testing unified X-ray/ultraviolet absorber models with NGC~5548', 
ApJ, 452, 230

\vskip 0.02 in 
\par\noindent\hangindent=10pt\hangafter=1
Murray N., Chiang J., Grossman S.A., Voit G.M., 
1995,
`Accretion disk winds from active galactic nuclei', 
ApJ, 451, 498

\vskip 0.02 in 
\par\noindent\hangindent=10pt\hangafter=1
Netzer H., 
1996, 
`X-ray lines in active galactic nuclei and photoionized gases', 
ApJ, 473, 781

\vskip 0.02 in 
\par\noindent\hangindent=10pt\hangafter=1
Netzer H., 
2001, 
`Physical processes in starburst and active galaxies', 
in Aretxaga I., Kunth D., M\'ujica R., eds, 
Advanced Lectures on the Starburst-AGN Connection. 
World Scientific, Singapore, p.~117

\vskip 0.02 in 
\par\noindent\hangindent=10pt\hangafter=1
Netzer H., 
2002, 
`The density and location of the X-ray absorbing gas in NGC~3516',
ApJ, 571, 256

\vskip 0.02 in 
\par\noindent\hangindent=10pt\hangafter=1
Nicastro F., Fiore F., Perola G.C., Elvis M., 
1999, 
`Ionized absorbers in active galactic nuclei: The role of collisional 
ionization and time-evolving photoionization',
ApJ, 512, 184

\vskip 0.02 in 
\par\noindent\hangindent=10pt\hangafter=1
Proga D., Stone J.M., Kallman T.R., 
2000, 
`Dynamics of line-driven disk winds in active galactic nuclei', 
ApJ, 543, 686

\vskip 0.02 in 
\par\noindent\hangindent=10pt\hangafter=1
Reynolds C.S., 
1997, 
`An X-ray spectral study of 24 type~1 active galactic nuclei', 
MNRAS, 286, 513

\vskip 0.02 in 
\par\noindent\hangindent=10pt\hangafter=1
Sako M., et~al., 
2001, 
`Complex resonance absorption structure in the X-ray 
spectrum of IRAS~13349+2438',
A\&A, 365, L168

\vskip 0.02 in 
\par\noindent\hangindent=10pt\hangafter=1
Silk J., Rees M.J., 
1998, 
`Quasars and galaxy formation', 
A\&A, 331, L1

\vskip 0.02 in 
\par\noindent\hangindent=10pt\hangafter=1
Turnshek D.A., 
1988,
`BALQSOs: Observations, models, and implications for narrow absorption line systems',
in Blades J.C., Turnshek D.A., Norman C.A., eds, 
QSO absorption lines: Probing the Universe. 
Cambridge University Press, Cambridge, p.~17

\vskip 0.02 in 
\par\noindent\hangindent=10pt\hangafter=1
Voit G.M., Weymann R.J., Korista K.T., 
1993, 
`Low-ionization broad absorption lines in quasars', 
ApJ, 413, 95

\vskip 0.02 in 
\par\noindent\hangindent=10pt\hangafter=1
Wu K.K.S., Fabian A.C., Nulsen P.E.J., 
2000, 
`Non-gravitational heating in the hierarchical formation of X-ray clusters', 
MNRAS, 318, 889


\end{document}